\documentclass[sigconf,authorversion]{acmart}
\AtBeginDocument{%
  }

\copyrightyear{2025}
\acmYear{2025}
\setcopyright{acmlicensed}
\acmConference[RecSys '25]{Proceedings of the Nineteenth ACM Conference on Recommender Systems}{September 22--26, 2025}{Prague, Czech Republic}
\acmBooktitle{Proceedings of the Nineteenth ACM Conference on Recommender Systems (RecSys '25), September 22--26, 2025, Prague, Czech Republic}
\acmDOI{10.1145/3705328.3748033}
\acmISBN{979-8-4007-1364-4/2025/09}

\usepackage{multirow}
\usepackage{stfloats}
\begin{document}

\title{Correcting the LogQ Correction: Revisiting Sampled Softmax for Large-Scale Retrieval}
\author{Kirill Khrylchenko}
\email{elightelol@gmail.com}
\orcid{0009-0007-3640-8795}
\affiliation{%
  \institution{Yandex}
  \city{Moscow}
  \country{Russia}
}

\author{Vladimir Baikalov}
\email{deadinside@yandex-team.ru}
\orcid{0009-0009-4864-2305}
\affiliation{%
  \institution{Yandex}
  \city{Moscow}
  \country{Russia}
}

\author{Sergei Makeev}
\email{neuralsrg@gmail.com}
\orcid{0009-0003-5451-6475}
\affiliation{%
  \institution{Yandex}
  \city{Moscow}
  \country{Russia}
}

\author{Artem Matveev}
\email{matfu21@ya.ru}
\orcid{0009-0004-0271-221X}
\affiliation{%
 \institution{Yandex}
  \city{Moscow}
  \country{Russia}
 }
 
\author{Sergei Liamaev}
\email{lyamaev.sergei@yandex.ru}
\orcid{0009-0009-6316-1091}
\affiliation{%
  \institution{Yandex}
  \city{Moscow}
  \country{Russia}
}

\renewcommand{\shortauthors}{Khrylchenko et al.}

\begin{abstract}
Two-tower neural networks are a popular architecture for the retrieval stage in recommender systems. These models are typically trained with a softmax loss over the item catalog. However, in web-scale settings, the item catalog is often prohibitively large, making full softmax infeasible. A common solution is sampled softmax, which approximates the full softmax using a small number of sampled negatives.

One practical and widely adopted approach is to use in-batch negatives, where negatives are drawn from items in the current mini-batch. However, this introduces a bias: items that appear more frequently in the batch (i.e., popular items) are penalized more heavily.

To mitigate this issue, a popular industry technique known as logQ correction adjusts the logits during training by subtracting the log-probability of an item appearing in the batch. This correction is derived by analyzing the bias in the gradient and applying importance sampling, effectively twice, using the in-batch distribution as a proposal distribution. While this approach improves model quality, it does not fully eliminate the bias.

In this work, we revisit the derivation of logQ correction and show that it overlooks a subtle but important detail: the positive item in the denominator is not Monte Carlo-sampled --- it is always present with probability 1. We propose a refined correction formula that accounts for this. Notably, our loss introduces an interpretable sample weight that reflects the model’s uncertainty --- the probability of misclassification under the current parameters. We evaluate our method on both public and proprietary datasets, demonstrating consistent improvements over the standard logQ correction.
\end{abstract}

\ccsdesc[500]{Information systems~Recommender systems}
\keywords{Recommender systems; Information Retrieval; Neural Networks}


\maketitle

\section{Introduction}
Recommender systems are widely used to alleviate information overload across domains with large item catalogs. Modern systems typically adopt a multi-stage architecture: a retrieval stage selects a small set of relevant candidates (hundreds or thousands), followed by a ranking stage that scores them. Retrieval requires global comparison over all items and is commonly trained using a softmax loss over the catalog. However, at web scale, computing the full softmax is computationally prohibitive due to large matrix multiplications and the possible need to compute item embeddings dynamically, which makes calculations at each step infeasible.

A standard workaround is \emph{sampled softmax}, which samples negative items to calculate softmax. A particularly efficient negatives source is \emph{in-batch negatives}, where items from the current mini-batch are reused as negative samples. This not only avoids redundant computation when item embeddings are generated dynamically, but also provides harder, more informative negatives compared to uniform sampling.

However, in-batch negatives introduce a bias: popular items tend to appear more frequently as negatives simply because they are more likely to occur in the training batches. As a result, they become over-represented in the negative set and may be excessively penalized during training, which can degrade the model’s ability to rank them correctly. A widely used fix is the \emph{logQ correction}\,\cite{logq}, which applies importance sampling to adjust for the in-batch distribution.

In this work, we revisit the derivation of logQ correction and reveal a subtle but important inconsistency. The standard derivation implicitly assumes that the positive item is sampled from the same proposal distribution as the negatives. In reality, however, the positive item is inserted deterministically and is not subject to sampling. This incorrect treatment of the positive item introduces systematic bias.

We address this inconsistency by introducing a corrected formulation of logQ correction. \newpage

\textbf{Our contributions} are as follows:
\begin{itemize}
\item We derive an improved version of logQ correction that explicitly accounts for the fact that positive item is deterministic and is not sampled. Our formulation naturally introduces a sample weighting scheme that reflects the model’s confidence, defined as the probability of misclassifying the positive item.
\item We demonstrate consistent improvements over standard logQ correction across public and proprietary datasets.
\item We bridge the gap between academic research and industrial practice by introducing logQ correction baselines on public sequential recommendation benchmarks.
\end{itemize}

\section{Related Work}

The use of softmax over large output spaces dates back to \textbf{neural probabilistic language models} of \citet{nplm}, where it was employed for the next-word prediction task. Due to the computational cost of summing over the entire vocabulary, \citet{bengio03} proposed sampling a small subset of candidates, typically from the unigram distribution. They also showed that this introduces biased gradient estimates, and suggested applying \emph{weighted importance sampling} to partially correct for the resulting bias. The subsequent shift toward subword modeling\,\cite{wordpiece, bert} in natural language processing substantially reduced vocabulary sizes, making full softmax tractable.

In contrast, SAR (Search, Ads, Recommendations) retrieval systems continue to operate at web scale, where sampled softmax remains essential due to the size of item and document catalogs, often reaching millions of entries. Early \textbf{two-tower models}, such as DSSM\,\cite{dssm} for query-document matching and YouTubeDNN\,\cite{youtubednn} for user-item recommendation, both relied on sampled softmax to scale training. YouTubeDNN also briefly discussed applying importance weighting to correct the bias introduced by non-uniform negative sampling. Later, \citet{logq} adopted the importance sampling approach from \citet{bengio03} to the setting of recommendation, leading to improved offline and online performance. \citet{mns} extended this line of work by combining in-batch negatives with uniformly sampled ones. Since then, logQ correction has become a standard practice in industrial two-tower retrieval models at companies like Google\,\cite{logq, mns}, Pinterest\,\cite{pinnerformer}, ByteDance\,\cite{trinity}, and Kuaishou\,\cite{Kuaishou}.

\textbf{Sequential recommenders} can often be viewed as a special case of retrieval two-tower models where user representations are derived from interaction sequences. They are typically trained to predict the next item. GRU4Rec\,\cite{gru4rec} used the BPR\,\cite{bpr} loss with in-batch negatives, while CASER\,\cite{caser} and SASRec\,\cite{sasrec} used binary cross-entropy (BCE) with three and one uniformly sampled negatives, respectively. BERT4Rec\,\cite{bert4rec} introduced a masked item prediction objective and initially outperformed SASRec\,\cite{sasrec}, but \citet{dross} later showed this was mainly due to the use of full softmax; \citet{pinnerformer} further confirmed that full softmax substantially improves SASRec itself in an industrial setting, using real-world user traffic at Pinterest. \citet{sampled_softmax} showed that sampled softmax outperforms traditional BCE and BPR losses, while \citet{gsasrec} proposed a confidence correction for BCE loss combined with increased negative sampling, achieving results comparable to or exceeding those of full softmax models.

Despite the widespread adoption of logQ correction in industry, academic work has largely overlooked this technique. Moreover, even in industry applications, the theoretical foundations of logQ correction remain imperfect. In this paper, we revisit logQ correction from a theoretical perspective, propose an improved formulation, and adapt it to academic benchmarks. We further evaluate its effectiveness by comparing against strong baselines such as gSASRec\,\cite{gsasrec}.

\section{Method}
\subsection{Problem Setup}

In large-scale recommendation systems, the retrieval stage aims to select a small subset of relevant items from an extremely large catalog. A common architecture for retrieval is the \textbf{two-tower model}, where user $u$ and item $d$ representations are produced independently, and their interaction is modeled via a simple dot product:
\[
f_\theta(u, d) = \langle g_\theta(u), h_\theta(d) \rangle,
\]
where $\theta$ denotes the model parameters, $g_\theta(u)$ is the user encoder, and $h_\theta(d)$ is the item encoder.

This setup can be viewed as an instance of \textbf{extreme multiclass classification}, where each class corresponds to a distinct item from a massive catalog.

The probability of a user $u$ interacting with an item $d$ is modeled using a softmax over the entire item catalog $\mathcal{D}$:
\[
P_\theta(d \mid u) = \frac{e^{f_\theta(u, d)}}{\sum_{d' \in \mathcal{D}} e^{f_\theta(u, d')}}.
\]

The associated loss function for an observed positive pair $(u, p)$ is the negative log-likelihood:
\[
\mathcal{L}_{\text{softmax}}(u, p) = -\log P_\theta(p \mid u) = -\log \frac{e^{f_\theta(u, p)}}{\sum_{d' \in \mathcal{D}} e^{f_\theta(u, d')}}.
\]

\subsection{LogQ Correction} \label{sec:sampled}
Since computing the full softmax over the entire item catalog is \emph{computationally infeasible} for large item catalogs, it is commonly approximated using sampled softmax:
\[
\mathcal{L}_{\text{sampled}}(u, p) = -\log \frac{e^{f_\theta(u, p)}}{e^{f_\theta(u, p)} + \sum_{i=1}^n e^{f_\theta(u, d_i)}},
\]
where ${d_1, \ldots, d_n}$ are negative items sampled from some distribution $Q(d)$ over the item catalog $\mathcal{D}$.

A variety of negative sampling strategies are used as $Q(d)$ in practice. One particularly popular choice is in-batch negative sampling, where positive items from other examples in the same training batch are reused as negatives. While this greatly reduces computation, it introduces a \emph{bias}: the resulting sampling distribution reflects the empirical popularity of items (i.e., a unigram distribution), leading to over-representation and over-penalization of popular items.

To understand this effect more formally, consider the gradient of the original softmax loss:
\[
\nabla_\theta \mathcal{L}_{\text{softmax}}(u, p) = -\nabla_\theta f_\theta(u, p) + \mathbb{E}_{d \sim P_\theta(d \mid u)} \left[\nabla_\theta f_\theta(u, d) \right],
\]
where the first term corresponds to the observed positive interaction, and the second term is an expectation over the entire item catalog under the model’s predicted distribution $P_\theta(d \mid u)$.

The use of a sampled approximation instead of the full softmax effectively shifts the expectation distribution from $P_\theta$ to a proposal distribution $Q$. This change introduces bias, particularly when $Q(d)$ significantly differs from $P_\theta(d \mid u)$.

To address this \emph{distribution mismatch}, \citet{bengio03} applied weighted importance sampling (also known as ``biased importance sampling''\,\cite{bengio03}), which adjusts both the expectation and the normalization by reweighting samples from $Q(d)$. This yields the standard logQ-corrected loss:
\[
\mathcal{L}_{\text{logQ}}(u, p) = -\log \frac{e^{f_\theta(u, p)}}{e^{f_\theta(u, p) - \log Q(p)} + \sum_{i=1}^n e^{f_\theta(u, d_i) - \log Q(d_i)}}.
\]

Note that the logQ correction for the positive item in the numerator, as done in the original work\,\cite{logq}, is unnecessary since it does not affect the gradient. However, removing it from the denominator~---~as done by \citet{mns}~---~does not follow directly from the \citet{bengio03} derivation, and we are unaware of a principled explanation for this variant.

\subsection{Correcting the Correction}
In the standard logQ correction, positive item $p$ is used together with sampled items to produce Monte Carlo estimate of the expectation over the full catalog. However, in contrast to negatives, it is not sampled from the proposal distribution $Q(d)$ --- for a given training instance, it is deterministically present with probability 1. This mismatch introduces additional bias.

To derive a more accurate gradient that accounts for the fact that positive is not sampled, let us first decompose the gradient of the full softmax loss:
\begin{align*} 
&\nabla_\theta \mathcal{L}_{\text{softmax}}(u, p) = \nabla_\theta \left[ -f_\theta(u, p) + \log \sum_{d \in \mathcal{D}} e^{f_\theta(u, d)} \right] \\
&= -\nabla_\theta f_\theta(u, p) + \frac{1}{\sum_{d \in \mathcal{D}} e^{f_\theta(u, d)}} \sum_{d \in \mathcal{D}} e^{f_\theta(u, d)} \nabla_\theta f_\theta(u, d)\\ 
&= -\nabla_\theta f_\theta(u, p) + \sum_{d \in \mathcal{D}} P_\theta(d \mid u) \nabla_\theta f_\theta(u, d) \\ 
&= -(1 - P_\theta(p \mid u)) \nabla_\theta f_\theta(u, p) + \sum_{d \in \mathcal{D} \setminus \{p\}} P_\theta(d \mid u) \nabla_\theta f_\theta(u, d) \\
&= (1 - P_\theta(p \mid u)) \left( -\nabla_\theta f_\theta(u, p) + \mathbb{E}_{d \sim P_\theta(d \mid u, d \neq p)} \left[ \nabla_\theta f_\theta(u, d) \right] \right),
\end{align*}
where in the last step we used the identity
\[
P_\theta(d \mid u) = (1 - P_\theta(p \mid u)) \cdot P_\theta(d \mid u, d \neq p), \quad \forall d \in \mathcal{D} \setminus \{p\}.
\]

To approximate the expectation, we apply weighted importance sampling:
\begin{align*} 
&\mathbb{E}_{d \sim P_\theta(d \mid u, d \neq p)} \left[ \nabla_\theta f_\theta(u, d) \right] \approx \\
&\approx \sum_{i=1}^n \frac{e^{f_\theta(u, d_i) - \log Q'(d_i)}}{\sum_{k=1}^n e^{f_\theta(u, d_k) - \log Q'(d_k)}} \nabla_\theta f_\theta(u, d_i), 
\end{align*}
where ${d_1, \ldots, d_n}$ are sampled from $Q'(d)$ --- proposal distribution over $\mathcal{D} \setminus \{p\}$.

The resulting gradient corresponds to minimizing the following weighted loss:
\[ \mathcal{L}_{\text{ours}}(u, p) = - w_{up} \log \frac{e^{f_\theta (u, p)}}{\sum_{i=1}^n e^{f_\theta (u, d_i) - \log Q'(d_i)}},
\]
where $w_{up} = \text{sg}(1 - P_\theta(p \mid u))$, $\text{sg}$ denotes the stop-gradient operation (no gradient flows through $w_{up}$), and $d_1, \dots, d_n$ are sampled from $Q'(d)$. Notably, the positive item is no longer present in softmax denominator.

The introduced weight $w_{up}$ reflects the model's confidence in distinguishing the positive item $p$ from the negatives: if the model assigns a high score to $p$ relative to the negatives, the weight will be small.

To estimate $P_\theta(p \mid u)$ for $w_{up}$, importance sampling is applied once more:
\begin{align*} 
P_\theta(p \mid u) 
&= \frac{e^{f_\theta(u, p)}}{\sum_{d \in \mathcal{D}} e^{f_\theta(u, d)}} = \frac{e^{f_\theta(u, p)}}{e^{f_\theta(u, p)} + \sum_{d \in \mathcal{D} \setminus \{p\}} e^{f_\theta(u, d)}} \\ 
&= \frac{e^{f_\theta(u, p)}}{e^{f_\theta(u, p)} + (|\mathcal{D}|-1) \mathbb{E}_{d \sim \text{Unif}(\mathcal{D} \setminus \{p\})} \left[ e^{f\theta(u, d)} \right]} \\ 
&\approx \frac{e^{f_\theta(u, p)}}{e^{f_\theta(u, p)} + \frac{1}{n} \sum_{i=1}^n e^{f_\theta(u, d_i) - \log Q'(d_i)}}, \quad d_i \sim Q'(d). 
\end{align*}
In practice, we reuse the same set of sampled negatives for both the loss computation and the calculation of sample weights $w_{up}$, ensuring computational efficiency.

\section{Experiments}

We evaluate our method in both academic and industrial setups to demonstrate its robustness across different environments. The industrial setup is based on a large-scale production system.

\paragraph{Negative sampling and logQ correction.}
In all experiments, we follow standard practice and draw negative items from a proposal distribution $Q$ --- via in-batch sampling, uniform sampling, or their mixture. To obtain the corrected distribution $Q'$, which excludes the positive item $p$ from sampling, we adopt a simple but effective approach: we sample negatives from $Q$ and discard the positive item if it appears.

Along with this masking, we replace $\log Q(d)$ with $\log Q'(d)$ in the loss. The exact computation of $Q'(d)$ depends on the setting:
\begin{itemize}
    \item In the \textbf{industrial setup}, computing exact $Q'(d)$ is infeasible due to the streaming nature of the data and the large item catalog. Instead, we approximate item frequencies using a \emph{count-min sketch}\,\cite{countminsketch}, which offers a memory-efficient way to estimate $Q(d)$. We empirically find that using $\log Q(d)$ as a proxy for $\log Q'(d)$ introduces negligible differences in performance, and adopt this approximation in all industrial experiments.
    \item In the \textbf{academic setup}, full access to the dataset allows computing $Q(d)$ exactly. For example, under in-batch sampling, $Q(d)$ corresponds to item frequencies: \begin{equation*}
        Q(d) = \frac{\#d}{N}, \qquad Q'(d) = \frac{\#d}{N - \#p} = \frac{N}{N - \#p} Q(d),
    \end{equation*}
    where $\#d$ is the number of times item $d$ appears in the dataset, and $N$ is the total number of item interactions.
\end{itemize}

\paragraph{Deduplication and negative diversity.}
To ensure a diverse set of negatives, we apply \emph{deduplication} to the in-batch item pool before sampling. It is especially important when using small batch sizes or when many sequences share the same items. 

\paragraph{Computational efficiency.}
Our improved logQ correction method introduces only a negligible computational overhead compared to the baseline. It requires a minor adjustment in the loss computation and does not require any changes to the model architecture or data pipeline.

\subsection{Academic Setup}

We conduct experiments on two public datasets: MovieLens-1M\,\cite{movielens} and Steam\,\cite{steam}, following the protocol of gSASRec\,\cite{gsasrec}. We adopt the \emph{leave-one-out evaluation} scheme, where the last user interaction is held out for testing and the second-to-last for validation. During evaluation, all non-positive items serve as negatives.

We also introduce \emph{temporal evaluation}\,\cite{timesplit1, timesplit2, timesplit3}, where all interactions are globally sorted by timestamp across the entire dataset, and the last $n$\% of interactions form the test set, with the preceding $n$\% used for validation ($n = 10$ for MovieLens-1M, $n = 1$ for Steam). This split is performed \emph{globally}, not per user, ensuring that the model is evaluated on future data relative to the training period.

Although the gSASRec protocol includes the Gowalla dataset\,\cite{gowalla}, we were unable to reproduce its reported results using the public implementation. We omit Gowalla from our main evaluation and leave a more thorough investigation for future work.

We report \textbf{Recall@20} and \textbf{NDCG@20}, averaging metrics across test interactions in the temporal split. All results are averaged over 5 independent runs.

As the backbone model, we use \textbf{SASRec}\,\cite{sasrec}, and compare the following training losses:
\begin{itemize}
    \item \textbf{Binary Cross-Entropy (BCE)} with a single uniformly sampled negative --- the original setup used in SASRec\,\cite{sasrec}. This approach suffers from two key limitations: the suboptimality of the BCE loss itself (softmax-based losses are often superior\,\cite{dross, pinnerformer}) and the minimal number of sampled negatives\,\cite{gsasrec}.
    \item \textbf{Full Softmax Loss}, which was a major advantage\,\cite{dross} of \\ BERT4Rec\,\cite{bert4rec} over SASRec. This theoretically ideal loss computes the normalization over the entire item catalog, but becomes computationally infeasible in real-world recommendation settings with large catalogs.
    \item \textbf{Sampled Softmax Loss}, discussed in detail in Section~\ref{sec:sampled}. Instead of summing over all items, it uses a subset of sampled negatives. We explore different strategies for sampling:
    \begin{itemize}
        \item \textbf{Uniform negatives}, drawn uniformly from the entire item catalog. These negatives tend to be less informative, and a large number is often needed to produce meaningful gradients.
        \item \textbf{In-batch negatives}, implicitly sampled from the unigram distribution. These are typically harder than uniform negatives and computationally cheaper, as they reuse items from the current mini-batch.
         \item \textbf{Mixed negative sampling (MNS)}\,\cite{mns}, which combines both sources above.
    \end{itemize}
    \item For sampled softmax, we compare three \textbf{logQ correction} strategies: no correction, the standard logQ correction\,\cite{logq}, and our proposed improved variant.
    \item \textbf{Generalized Binary Cross-Entropy (gBCE)}, proposed in gSASRec\,\cite{gsasrec}, which improves upon BCE by using a generalized sigmoid function for positive samples and by training with a larger pool of negative samples.
\end{itemize}

\begin{paragraph}{Implementation Details}
We release our code\footnote{\url{https://github.com/NonameUntitled/logq}} based on the official pytorch gSASRec\footnote{\url{https://github.com/asash/gSASRec-pytorch}} repository. Unless stated otherwise, we follow its default hyperparameters.

For gBCE and sampled softmax, we sample 256 negatives. In the MNS setup, 128 negatives are sampled uniformly and 128 from in-batch items. Importantly, for mixed sampling, $\log Q$ values are based on unigram frequencies and not adjusted for the uniform portion.

\end{paragraph}

\subsection{Industrial Setup} \label{sec:ind}

We further validate our method on a \textbf{large-scale retrieval task} using an in-house production dataset with over 300 billion user-item interactions collected over a year on a major music streaming platform.

\paragraph{Model Architecture and Training.} Our production model follows a \textbf{two-tower architecture}, trained to predict future positive engagements based on user interaction histories. During training, we use sampled softmax with mixed negative sampling\,\cite{mns} and logQ correction\,\cite{logq}, combining 8192 uniformly sampled negatives with 8192 unique in-batch negatives.

Due to confidentiality, we do not disclose dataset details or the exact model architecture.

\paragraph{Evaluation Protocol.} For evaluation, we use user-item interactions from the day following the training period, measuring performance separately for consumption and engagement signals. The retrieval index includes the top 2 million most popular items from training, along with all positive user-item interactions from the test day. User embeddings are computed from data preceding the test day. We report \textbf{Recall@k} for $k \in \{10, 100, 1000\}$. The goal is to retrieve, for each user, as many of their next-day positive interactions as possible within a relatively small candidate set.

\paragraph{Why Recall@k?} In our industrial setup, the model serves as a \emph{retrieval model}, whose primary goal is to produce a large pool of relevant items (typically thousands) for downstream reranking. Consequently, recall is the key metric: improvements in Recall@100 or Recall@1000 translate directly into measurable gains in downstream system performance. In contrast, metrics such as NDCG, while informative for final ranking tasks, are less meaningful in the retrieval stage and can be misleading when a separate model handles ranking. This practice is common across large-scale retrieval systems in industry\,\cite{logq, mns, taobao, pinnerformer, itemsage, amazon}.

For public datasets, we additionally report NDCG to align with academic conventions. 

\paragraph{Index Size in Offline Evaluation.} While production indices often contain tens or hundreds of millions of items, offline evaluation is typically performed on a smaller sampled subset (e.g., 1~--~10M candidates). Our use of a 2M-item index follows this standard practice, as seen in prior work from Pinterest\,\cite{pinnerformer, itemsage} and Amazon\,\cite{amazon}.

\subsection{Results}

\begin{paragraph}{Academic Setup}

\begin{table*}[htbp!]
    \caption{Recall@20 and NDCG@20 performance for different models comparing leave-one-out and temporal split validation on the public datasets}
    \label{tab:academic}
    \centering
    \setlength{\tabcolsep}{2.0pt}
    \begin{tabular}{lcccccccc}
      \toprule
      \multicolumn{1}{c}{\multirow{3}{*}{\textbf{Loss}}} & \multicolumn{4}{c}{\textbf{Leave-One-Out}} & \multicolumn{4}{c}{\textbf{Temporal Split}} \\
      \cmidrule(lr){2-5} \cmidrule(lr){6-9}
      & \multicolumn{2}{c}{\textit{MovieLens-1M}} & \multicolumn{2}{c}{\textit{Steam}} & \multicolumn{2}{c}{\textit{MovieLens-1M}} & \multicolumn{2}{c}{\textit{Steam}} \\
      & NDCG@20 & R@20 & NDCG@20 & R@20 & NDCG@20 & R@20 & NDCG@20 & R@20 \\
      \midrule
      \textbf{BCE} & 0.1564 & 0.3468 & 0.0640 & 0.1540 & 0.0997 & 0.2318 & 0.0663 & 0.1472 \\[0.1cm]
      \textbf{gBCE} & 0.1964 & \textbf{0.3949} & \underline{0.0813} & \underline{0.1844} & 0.1210 & 0.2632 & \underline{0.0777} & \underline{0.1701} \\[0.1cm]
      \textbf{Sampled Softmax} &  & & & & & & & \\
      \hspace*{0.3cm} \textit{Uniform negatives}
        & 0.1921 & 0.3889 & 0.0811 & 0.1842 & 0.1186 & 0.2609 & \textbf{0.0801} & \textbf{0.1768} \\
      \hspace*{0.3cm} \textit{In-batch negatives} & & & & & & & & \\
      \hspace*{0.7cm} \normalfont{Without logQ}
        & 0.1868 & 0.3838 & 0.0394 & 0.0953 & 0.1117 & 0.2481 & 0.0432 & 0.0990 \\
      \hspace*{0.7cm} \normalfont{With standard logQ}
        & 0.1942 & 0.3889 & 0.0727 & 0.1698 & 0.1270 & 0.2762 & 0.0632 & 0.1483 \\
      \hspace*{0.7cm} \normalfont{With improved logQ}
        & \underline{0.1969} & \underline{0.3947} & 0.0719 & 0.1689 & 0.1269 & 0.2766 & 0.0693 & 0.1588 \\[0.1cm]
      \hspace*{0.3cm} \textit{Mixed negatives} & & & & & & & & \\
      \hspace*{0.7cm} \normalfont{Without logQ}
        & 0.1882 & 0.3853 & 0.0645 & 0.1470 & 0.1130 & 0.2514 & 0.0656 & 0.1389 \\
      \hspace*{0.7cm} \normalfont{With standard logQ}
        & 0.1952 & 0.3893 & 0.0729 & 0.1694 & \textbf{0.1292} & \textbf{0.2800} & 0.0639 & 0.1485 \\
      \hspace*{0.7cm} \normalfont{With improved logQ}
        & \textbf{0.1972} & 0.3937 & 0.0748 & 0.1730 & \underline{0.1281} & \underline{0.2792} & 0.0701 & 0.1609 \\[0.1cm]
      \textbf{Full Softmax}
        & 0.1937 & 0.3911 & \textbf{0.0822} & \textbf{0.1856} & 0.1205 & 0.2640 & 0.0770 & 0.1688 \\
      \bottomrule
    \end{tabular}
\end{table*}
Table~\ref{tab:academic} shows that \emph{no single loss dominates} across all datasets and evaluation setups. Full softmax performs best only on Steam (leave-one-out), while uniform sampled softmax unexpectedly leads on Steam (temporal split). On MovieLens, logQ corrections achieve the strongest performance under temporal-split evaluation, and are competitive under leave-one-out as well --- falling slightly behind gBCE in Recall@20 but outperforming it in NDCG@20, with minimal margin.

In-batch negatives without any correction result in substantially degraded metrics. Applying logQ correction recovers most of the gap, and combining it with MNS further improves performance --- often pushing logQ-corrected models to the top. Our improved logQ correction with MNS \emph{yields consistent gains over the standard variant}: the improvement is large on MovieLens (leave-one-out) and across all metrics on Steam, while slightly trailing standard logQ only on MovieLens (temporal split).

Finally, the relative effectiveness of different training objectives varies noticeably between leave-one-out and temporal splits. This suggests that conclusions drawn from leave-one-out evaluation may not transfer well to more realistic, time-aware protocols.

\end{paragraph}

\begin{paragraph}{Industrial Setup}

\begin{table}[t]
\centering
\caption{Comparison of logQ correction techniques on the production setup, Recall@k}
\label{tab:industry}
\small
\setlength{\tabcolsep}{4pt}
\begin{tabular}{lcccccc}
\toprule
\multirow{2}{*}{\parbox{1.4cm}{\begin{flushleft} \textbf{LogQ Correction} \end{flushleft}}} & \multicolumn{3}{c}{\textbf{Consumption}} & \multicolumn{3}{c}{\textbf{Engagement}} \\
\cmidrule(lr){2-4} \cmidrule(lr){5-7}
 & @10 & @100 & @1000 & @10 & @100 & @1000 \\
\midrule
Without &  0.0679 & 0.0866 & 0.3133  & 0.0280 & 0.0990 & 0.2992 \\
Standard & \textbf{0.0756} & 0.1140 & 0.4308 & \textbf{0.0304} & 0.1211 & 0.4036 \\
Improved & 0.0700 & \textbf{0.1157} & \textbf{0.4616}  & 0.0279 & \textbf{0.1222} & \textbf{0.4345}  \\
\bottomrule
\end{tabular}
\end{table}

Table~\ref{tab:industry} reports Recall@k on our large-scale production setup. As discussed in the section \ref{sec:ind}, high values of $k$ are most relevant in retrieval scenarios, where the goal is to surface a large candidate pool for downstream ranking. On this axis, both standard and improved logQ corrections yield substantial gains over the uncorrected baseline, with the improved variant showing a growing advantage as $k$ increases.

Interestingly, at $k = 10$, the improved method performs on par with the uncorrected model, while the standard version achieves slightly better results. Although this regime is less critical for our use case, it suggests there may be subtle differences in how the corrections affect the head versus the tail of the retrieved list. We leave a deeper investigation of this phenomenon to future work.
\end{paragraph}

\section{Conclusion}
In this work, we revisited logQ correction for sampled softmax training in large-scale retrieval systems. We identified limitations of the standard correction when used with in-batch negatives and proposed an improved variant that better aligns training with the ideal softmax objective. Through extensive evaluation on public benchmarks and a large-scale industrial dataset, we demonstrated that our correction consistently improves retrieval quality over standard logQ correction baselines. Furthermore, we highlighted the importance of evaluation setup and negative sampling strategy, showing that our method is robust across validation schemes and scales effectively to production settings.

\bibliographystyle{ACM-Reference-Format}
\bibliography{sample-base}


\begin{thebibliography}{29}


\ifx \showCODEN    \undefined \def \showCODEN     #1{\unskip}     \fi
\ifx \showISBNx    \undefined \def \showISBNx     #1{\unskip}     \fi
\ifx \showISBNxiii \undefined \def \showISBNxiii  #1{\unskip}     \fi
\ifx \showISSN     \undefined \def \showISSN      #1{\unskip}     \fi
\ifx \showLCCN     \undefined \def \showLCCN      #1{\unskip}     \fi
\ifx \shownote     \undefined \def \shownote      #1{#1}          \fi
\ifx \showarticletitle \undefined \def \showarticletitle #1{#1}   \fi
\ifx \showURL      \undefined \def \showURL       {\relax}        \fi
\providecommand\bibfield[2]{#2}
\providecommand\bibinfo[2]{#2}
\providecommand\natexlab[1]{#1}
\providecommand\showeprint[2][]{arXiv:#2}

\bibitem[Baltescu et~al\mbox{.}(2022)]%
        {itemsage}
\bibfield{author}{\bibinfo{person}{Paul Baltescu}, \bibinfo{person}{Haoyu Chen}, \bibinfo{person}{Nikil Pancha}, \bibinfo{person}{Andrew Zhai}, \bibinfo{person}{Jure Leskovec}, {and} \bibinfo{person}{Charles Rosenberg}.} \bibinfo{year}{2022}\natexlab{}.
\newblock \showarticletitle{ItemSage: Learning Product Embeddings for Shopping Recommendations at Pinterest}. In \bibinfo{booktitle}{\emph{Proceedings of the 28th ACM SIGKDD Conference on Knowledge Discovery and Data Mining}} (Washington DC, USA) \emph{(\bibinfo{series}{KDD '22})}. \bibinfo{publisher}{Association for Computing Machinery}, \bibinfo{address}{New York, NY, USA}, \bibinfo{pages}{2703–2711}.
\newblock
\showISBNx{9781450393850}
\href{https://doi.org/10.1145/3534678.3539170}{doi:\nolinkurl{10.1145/3534678.3539170}}


\bibitem[Bengio et~al\mbox{.}(2003)]%
        {nplm}
\bibfield{author}{\bibinfo{person}{Yoshua Bengio}, \bibinfo{person}{R\'{e}jean Ducharme}, \bibinfo{person}{Pascal Vincent}, {and} \bibinfo{person}{Christian Janvin}.} \bibinfo{year}{2003}\natexlab{}.
\newblock \showarticletitle{A neural probabilistic language model}.
\newblock \bibinfo{journal}{\emph{J. Mach. Learn. Res.}} \bibinfo{volume}{3}, \bibinfo{number}{null} (\bibinfo{date}{March} \bibinfo{year}{2003}), \bibinfo{pages}{1137–1155}.
\newblock
\showISSN{1532-4435}


\bibitem[Bengio and Senecal(2003)]%
        {bengio03}
\bibfield{author}{\bibinfo{person}{Yoshua Bengio} {and} \bibinfo{person}{Jean-S{\'{e}}bastien Senecal}.} \bibinfo{year}{2003}\natexlab{}.
\newblock \showarticletitle{Quick Training of Probabilistic Neural Nets by Importance Sampling}. In \bibinfo{booktitle}{\emph{Proceedings of the Ninth International Workshop on Artificial Intelligence and Statistics}} \emph{(\bibinfo{series}{Proceedings of Machine Learning Research}, Vol.~\bibinfo{volume}{R4})}, \bibfield{editor}{\bibinfo{person}{Christopher~M. Bishop} {and} \bibinfo{person}{Brendan~J. Frey}} (Eds.). \bibinfo{publisher}{PMLR}, \bibinfo{pages}{17--24}.
\newblock
\urldef\tempurl%
\url{https://proceedings.mlr.press/r4/bengio03a.html}
\showURL{%
\tempurl}
\newblock
\shownote{Reissued by PMLR on 01 April 2021.}.


\bibitem[Cho et~al\mbox{.}(2011)]%
        {gowalla}
\bibfield{author}{\bibinfo{person}{Eunjoon Cho}, \bibinfo{person}{Seth~A. Myers}, {and} \bibinfo{person}{Jure Leskovec}.} \bibinfo{year}{2011}\natexlab{}.
\newblock \showarticletitle{Friendship and mobility: user movement in location-based social networks}. In \bibinfo{booktitle}{\emph{Proceedings of the 17th ACM SIGKDD International Conference on Knowledge Discovery and Data Mining}} (San Diego, California, USA) \emph{(\bibinfo{series}{KDD '11})}. \bibinfo{publisher}{Association for Computing Machinery}, \bibinfo{address}{New York, NY, USA}, \bibinfo{pages}{1082–1090}.
\newblock
\showISBNx{9781450308137}
\href{https://doi.org/10.1145/2020408.2020579}{doi:\nolinkurl{10.1145/2020408.2020579}}


\bibitem[Cormode and Muthukrishnan(2005)]%
        {countminsketch}
\bibfield{author}{\bibinfo{person}{Graham Cormode} {and} \bibinfo{person}{S. Muthukrishnan}.} \bibinfo{year}{2005}\natexlab{}.
\newblock \showarticletitle{An improved data stream summary: the count-min sketch and its applications}.
\newblock \bibinfo{journal}{\emph{Journal of Algorithms}} \bibinfo{volume}{55}, \bibinfo{number}{1} (\bibinfo{year}{2005}), \bibinfo{pages}{58--75}.
\newblock
\showISSN{0196-6774}
\href{https://doi.org/10.1016/j.jalgor.2003.12.001}{doi:\nolinkurl{10.1016/j.jalgor.2003.12.001}}


\bibitem[Covington et~al\mbox{.}(2016)]%
        {youtubednn}
\bibfield{author}{\bibinfo{person}{Paul Covington}, \bibinfo{person}{Jay Adams}, {and} \bibinfo{person}{Emre Sargin}.} \bibinfo{year}{2016}\natexlab{}.
\newblock \showarticletitle{Deep Neural Networks for YouTube Recommendations}. In \bibinfo{booktitle}{\emph{Proceedings of the 10th ACM Conference on Recommender Systems}}. \bibinfo{address}{New York, NY, USA}.
\newblock


\bibitem[Devlin et~al\mbox{.}(2019)]%
        {bert}
\bibfield{author}{\bibinfo{person}{Jacob Devlin}, \bibinfo{person}{Ming-Wei Chang}, \bibinfo{person}{Kenton Lee}, {and} \bibinfo{person}{Kristina Toutanova}.} \bibinfo{year}{2019}\natexlab{}.
\newblock \showarticletitle{{BERT}: Pre-training of Deep Bidirectional Transformers for Language Understanding}. In \bibinfo{booktitle}{\emph{Proceedings of the 2019 Conference of the North {A}merican Chapter of the Association for Computational Linguistics: Human Language Technologies, Volume 1 (Long and Short Papers)}}, \bibfield{editor}{\bibinfo{person}{Jill Burstein}, \bibinfo{person}{Christy Doran}, {and} \bibinfo{person}{Thamar Solorio}} (Eds.). \bibinfo{publisher}{Association for Computational Linguistics}, \bibinfo{address}{Minneapolis, Minnesota}, \bibinfo{pages}{4171--4186}.
\newblock
\href{https://doi.org/10.18653/v1/N19-1423}{doi:\nolinkurl{10.18653/v1/N19-1423}}


\bibitem[Harper and Konstan(2015)]%
        {movielens}
\bibfield{author}{\bibinfo{person}{F.~Maxwell Harper} {and} \bibinfo{person}{Joseph~A. Konstan}.} \bibinfo{year}{2015}\natexlab{}.
\newblock \showarticletitle{The MovieLens Datasets: History and Context}.
\newblock \bibinfo{journal}{\emph{ACM Trans. Interact. Intell. Syst.}} \bibinfo{volume}{5}, \bibinfo{number}{4}, Article \bibinfo{articleno}{19} (\bibinfo{date}{Dec.} \bibinfo{year}{2015}), \bibinfo{numpages}{19}~pages.
\newblock
\showISSN{2160-6455}
\href{https://doi.org/10.1145/2827872}{doi:\nolinkurl{10.1145/2827872}}


\bibitem[Hidasi et~al\mbox{.}(2016)]%
        {gru4rec}
\bibfield{author}{\bibinfo{person}{Bal{\'{a}}zs Hidasi}, \bibinfo{person}{Alexandros Karatzoglou}, \bibinfo{person}{Linas Baltrunas}, {and} \bibinfo{person}{Domonkos Tikk}.} \bibinfo{year}{2016}\natexlab{}.
\newblock \showarticletitle{Session-based Recommendations with Recurrent Neural Networks}. In \bibinfo{booktitle}{\emph{4th International Conference on Learning Representations, {ICLR} 2016, San Juan, Puerto Rico, May 2-4, 2016, Conference Track Proceedings}}, \bibfield{editor}{\bibinfo{person}{Yoshua Bengio} {and} \bibinfo{person}{Yann LeCun}} (Eds.).
\newblock
\urldef\tempurl%
\url{http://arxiv.org/abs/1511.06939}
\showURL{%
\tempurl}


\bibitem[Huang et~al\mbox{.}(2013)]%
        {dssm}
\bibfield{author}{\bibinfo{person}{Po-Sen Huang}, \bibinfo{person}{Xiaodong He}, \bibinfo{person}{Jianfeng Gao}, \bibinfo{person}{Li Deng}, \bibinfo{person}{Alex Acero}, {and} \bibinfo{person}{Larry Heck}.} \bibinfo{year}{2013}\natexlab{}.
\newblock \showarticletitle{Learning deep structured semantic models for web search using clickthrough data}. In \bibinfo{booktitle}{\emph{Proceedings of the 22nd ACM International Conference on Information \& Knowledge Management}} (San Francisco, California, USA) \emph{(\bibinfo{series}{CIKM '13})}. \bibinfo{publisher}{Association for Computing Machinery}, \bibinfo{address}{New York, NY, USA}, \bibinfo{pages}{2333–2338}.
\newblock
\showISBNx{9781450322638}
\href{https://doi.org/10.1145/2505515.2505665}{doi:\nolinkurl{10.1145/2505515.2505665}}


\bibitem[Ji et~al\mbox{.}(2023)]%
        {timesplit2}
\bibfield{author}{\bibinfo{person}{Yitong Ji}, \bibinfo{person}{Aixin Sun}, \bibinfo{person}{Jie Zhang}, {and} \bibinfo{person}{Chenliang Li}.} \bibinfo{year}{2023}\natexlab{}.
\newblock \showarticletitle{A Critical Study on Data Leakage in Recommender System Offline Evaluation}.
\newblock \bibinfo{journal}{\emph{ACM Trans. Inf. Syst.}} \bibinfo{volume}{41}, \bibinfo{number}{3}, Article \bibinfo{articleno}{75} (\bibinfo{date}{Feb.} \bibinfo{year}{2023}), \bibinfo{numpages}{27}~pages.
\newblock
\showISSN{1046-8188}
\href{https://doi.org/10.1145/3569930}{doi:\nolinkurl{10.1145/3569930}}


\bibitem[Kang and McAuley(2018)]%
        {sasrec}
\bibfield{author}{\bibinfo{person}{Wang-Cheng Kang} {and} \bibinfo{person}{Julian McAuley}.} \bibinfo{year}{2018}\natexlab{}.
\newblock \showarticletitle{Self-Attentive Sequential Recommendation}. In \bibinfo{booktitle}{\emph{2018 IEEE International Conference on Data Mining (ICDM)}}. \bibinfo{pages}{197--206}.
\newblock
\href{https://doi.org/10.1109/ICDM.2018.00035}{doi:\nolinkurl{10.1109/ICDM.2018.00035}}


\bibitem[Klenitskiy and Vasilev(2023)]%
        {dross}
\bibfield{author}{\bibinfo{person}{Anton Klenitskiy} {and} \bibinfo{person}{Alexey Vasilev}.} \bibinfo{year}{2023}\natexlab{}.
\newblock \showarticletitle{Turning Dross Into Gold Loss: is BERT4Rec really better than SASRec?}. In \bibinfo{booktitle}{\emph{Proceedings of the 17th ACM Conference on Recommender Systems}} (Singapore, Singapore) \emph{(\bibinfo{series}{RecSys '23})}. \bibinfo{publisher}{Association for Computing Machinery}, \bibinfo{address}{New York, NY, USA}, \bibinfo{pages}{1120–1125}.
\newblock
\showISBNx{9798400702419}
\href{https://doi.org/10.1145/3604915.3610644}{doi:\nolinkurl{10.1145/3604915.3610644}}


\bibitem[Li et~al\mbox{.}(2021)]%
        {taobao}
\bibfield{author}{\bibinfo{person}{Sen Li}, \bibinfo{person}{Fuyu Lv}, \bibinfo{person}{Taiwei Jin}, \bibinfo{person}{Guli Lin}, \bibinfo{person}{Keping Yang}, \bibinfo{person}{Xiaoyi Zeng}, \bibinfo{person}{Xiao-Ming Wu}, {and} \bibinfo{person}{Qianli Ma}.} \bibinfo{year}{2021}\natexlab{}.
\newblock \showarticletitle{Embedding-based Product Retrieval in Taobao Search}. In \bibinfo{booktitle}{\emph{Proceedings of the 27th ACM SIGKDD Conference on Knowledge Discovery \& Data Mining}} (Virtual Event, Singapore) \emph{(\bibinfo{series}{KDD '21})}. \bibinfo{publisher}{Association for Computing Machinery}, \bibinfo{address}{New York, NY, USA}, \bibinfo{pages}{3181–3189}.
\newblock
\showISBNx{9781450383325}
\href{https://doi.org/10.1145/3447548.3467101}{doi:\nolinkurl{10.1145/3447548.3467101}}


\bibitem[Liu et~al\mbox{.}(2024)]%
        {Kuaishou}
\bibfield{author}{\bibinfo{person}{Chi Liu}, \bibinfo{person}{Jiangxia Cao}, \bibinfo{person}{Rui Huang}, \bibinfo{person}{Kai Zheng}, \bibinfo{person}{Qiang Luo}, \bibinfo{person}{Kun Gai}, {and} \bibinfo{person}{Guorui Zhou}.} \bibinfo{year}{2024}\natexlab{}.
\newblock \bibinfo{title}{KuaiFormer: Transformer-Based Retrieval at Kuaishou}.
\newblock
\showeprint[arxiv]{2411.10057}~[cs.IR]
\urldef\tempurl%
\url{https://arxiv.org/abs/2411.10057}
\showURL{%
\tempurl}


\bibitem[Meng et~al\mbox{.}(2020)]%
        {timesplit3}
\bibfield{author}{\bibinfo{person}{Zaiqiao Meng}, \bibinfo{person}{Richard McCreadie}, \bibinfo{person}{Craig Macdonald}, {and} \bibinfo{person}{Iadh Ounis}.} \bibinfo{year}{2020}\natexlab{}.
\newblock \showarticletitle{Exploring Data Splitting Strategies for the Evaluation of Recommendation Models}. In \bibinfo{booktitle}{\emph{Proceedings of the 14th ACM Conference on Recommender Systems}} (Virtual Event, Brazil) \emph{(\bibinfo{series}{RecSys '20})}. \bibinfo{publisher}{Association for Computing Machinery}, \bibinfo{address}{New York, NY, USA}, \bibinfo{pages}{681–686}.
\newblock
\showISBNx{9781450375832}
\href{https://doi.org/10.1145/3383313.3418479}{doi:\nolinkurl{10.1145/3383313.3418479}}


\bibitem[Nigam et~al\mbox{.}(2019)]%
        {amazon}
\bibfield{author}{\bibinfo{person}{Priyanka Nigam}, \bibinfo{person}{Yiwei Song}, \bibinfo{person}{Vijai Mohan}, \bibinfo{person}{Vihan Lakshman}, \bibinfo{person}{Weitian~(Allen) Ding}, \bibinfo{person}{Ankit Shingavi}, \bibinfo{person}{Choon~Hui Teo}, \bibinfo{person}{Hao Gu}, {and} \bibinfo{person}{Bing Yin}.} \bibinfo{year}{2019}\natexlab{}.
\newblock \showarticletitle{Semantic Product Search}. In \bibinfo{booktitle}{\emph{Proceedings of the 25th ACM SIGKDD International Conference on Knowledge Discovery \& Data Mining}} (Anchorage, AK, USA) \emph{(\bibinfo{series}{KDD '19})}. \bibinfo{publisher}{Association for Computing Machinery}, \bibinfo{address}{New York, NY, USA}, \bibinfo{pages}{2876–2885}.
\newblock
\showISBNx{9781450362016}
\href{https://doi.org/10.1145/3292500.3330759}{doi:\nolinkurl{10.1145/3292500.3330759}}


\bibitem[Pancha et~al\mbox{.}(2022)]%
        {pinnerformer}
\bibfield{author}{\bibinfo{person}{Nikil Pancha}, \bibinfo{person}{Andrew Zhai}, \bibinfo{person}{Jure Leskovec}, {and} \bibinfo{person}{Charles Rosenberg}.} \bibinfo{year}{2022}\natexlab{}.
\newblock \showarticletitle{PinnerFormer: Sequence Modeling for User Representation at Pinterest}. In \bibinfo{booktitle}{\emph{Proceedings of the 28th ACM SIGKDD Conference on Knowledge Discovery and Data Mining}} (Washington DC, USA) \emph{(\bibinfo{series}{KDD '22})}. \bibinfo{publisher}{Association for Computing Machinery}, \bibinfo{address}{New York, NY, USA}, \bibinfo{pages}{3702–3712}.
\newblock
\showISBNx{9781450393850}
\href{https://doi.org/10.1145/3534678.3539156}{doi:\nolinkurl{10.1145/3534678.3539156}}


\bibitem[Pathak et~al\mbox{.}(2017)]%
        {steam}
\bibfield{author}{\bibinfo{person}{Apurva Pathak}, \bibinfo{person}{Kshitiz Gupta}, {and} \bibinfo{person}{Julian McAuley}.} \bibinfo{year}{2017}\natexlab{}.
\newblock \showarticletitle{Generating and Personalizing Bundle Recommendations on Steam}. In \bibinfo{booktitle}{\emph{Proceedings of the 40th International ACM SIGIR Conference on Research and Development in Information Retrieval}} (Shinjuku, Tokyo, Japan) \emph{(\bibinfo{series}{SIGIR '17})}. \bibinfo{publisher}{Association for Computing Machinery}, \bibinfo{address}{New York, NY, USA}, \bibinfo{pages}{1073–1076}.
\newblock
\showISBNx{9781450350228}
\href{https://doi.org/10.1145/3077136.3080724}{doi:\nolinkurl{10.1145/3077136.3080724}}


\bibitem[Petrov and Macdonald(2023)]%
        {gsasrec}
\bibfield{author}{\bibinfo{person}{Aleksandr~Vladimirovich Petrov} {and} \bibinfo{person}{Craig Macdonald}.} \bibinfo{year}{2023}\natexlab{}.
\newblock \showarticletitle{gSASRec: Reducing Overconfidence in Sequential Recommendation Trained with Negative Sampling}. In \bibinfo{booktitle}{\emph{Proceedings of the 17th ACM Conference on Recommender Systems}} (Singapore, Singapore) \emph{(\bibinfo{series}{RecSys '23})}. \bibinfo{publisher}{Association for Computing Machinery}, \bibinfo{address}{New York, NY, USA}, \bibinfo{pages}{116–128}.
\newblock
\showISBNx{9798400702419}
\href{https://doi.org/10.1145/3604915.3608783}{doi:\nolinkurl{10.1145/3604915.3608783}}


\bibitem[Rendle et~al\mbox{.}(2009)]%
        {bpr}
\bibfield{author}{\bibinfo{person}{Steffen Rendle}, \bibinfo{person}{Christoph Freudenthaler}, \bibinfo{person}{Zeno Gantner}, {and} \bibinfo{person}{Lars Schmidt-Thieme}.} \bibinfo{year}{2009}\natexlab{}.
\newblock \showarticletitle{BPR: Bayesian personalized ranking from implicit feedback}. In \bibinfo{booktitle}{\emph{Proceedings of the Twenty-Fifth Conference on Uncertainty in Artificial Intelligence}} (Montreal, Quebec, Canada) \emph{(\bibinfo{series}{UAI '09})}. \bibinfo{publisher}{AUAI Press}, \bibinfo{address}{Arlington, Virginia, USA}, \bibinfo{pages}{452–461}.
\newblock
\showISBNx{9780974903958}


\bibitem[Sun(2023)]%
        {timesplit1}
\bibfield{author}{\bibinfo{person}{Aixin Sun}.} \bibinfo{year}{2023}\natexlab{}.
\newblock \showarticletitle{Take a Fresh Look at Recommender Systems from an Evaluation Standpoint}. In \bibinfo{booktitle}{\emph{Proceedings of the 46th International ACM SIGIR Conference on Research and Development in Information Retrieval}} (Taipei, Taiwan) \emph{(\bibinfo{series}{SIGIR '23})}. \bibinfo{publisher}{Association for Computing Machinery}, \bibinfo{address}{New York, NY, USA}, \bibinfo{pages}{2629–2638}.
\newblock
\showISBNx{9781450394086}
\href{https://doi.org/10.1145/3539618.3591931}{doi:\nolinkurl{10.1145/3539618.3591931}}


\bibitem[Sun et~al\mbox{.}(2019)]%
        {bert4rec}
\bibfield{author}{\bibinfo{person}{Fei Sun}, \bibinfo{person}{Jun Liu}, \bibinfo{person}{Jian Wu}, \bibinfo{person}{Changhua Pei}, \bibinfo{person}{Xiao Lin}, \bibinfo{person}{Wenwu Ou}, {and} \bibinfo{person}{Peng Jiang}.} \bibinfo{year}{2019}\natexlab{}.
\newblock \showarticletitle{BERT4Rec: Sequential Recommendation with Bidirectional Encoder Representations from Transformer}. In \bibinfo{booktitle}{\emph{Proceedings of the 28th ACM International Conference on Information and Knowledge Management}} (Beijing, China) \emph{(\bibinfo{series}{CIKM '19})}. \bibinfo{publisher}{ACM}, \bibinfo{address}{New York, NY, USA}, \bibinfo{pages}{1441--1450}.
\newblock
\showISBNx{978-1-4503-6976-3}
\href{https://doi.org/10.1145/3357384.3357895}{doi:\nolinkurl{10.1145/3357384.3357895}}


\bibitem[Tang and Wang(2018)]%
        {caser}
\bibfield{author}{\bibinfo{person}{Jiaxi Tang} {and} \bibinfo{person}{Ke Wang}.} \bibinfo{year}{2018}\natexlab{}.
\newblock \showarticletitle{Personalized Top-N Sequential Recommendation via Convolutional Sequence Embedding}. In \bibinfo{booktitle}{\emph{Proceedings of the Eleventh ACM International Conference on Web Search and Data Mining}} (Marina Del Rey, CA, USA) \emph{(\bibinfo{series}{WSDM '18})}. \bibinfo{publisher}{Association for Computing Machinery}, \bibinfo{address}{New York, NY, USA}, \bibinfo{pages}{565–573}.
\newblock
\showISBNx{9781450355810}
\href{https://doi.org/10.1145/3159652.3159656}{doi:\nolinkurl{10.1145/3159652.3159656}}


\bibitem[Wu et~al\mbox{.}(2024)]%
        {sampled_softmax}
\bibfield{author}{\bibinfo{person}{Jiancan Wu}, \bibinfo{person}{Xiang Wang}, \bibinfo{person}{Xingyu Gao}, \bibinfo{person}{Jiawei Chen}, \bibinfo{person}{Hongcheng Fu}, {and} \bibinfo{person}{Tianyu Qiu}.} \bibinfo{year}{2024}\natexlab{}.
\newblock \showarticletitle{On the Effectiveness of Sampled Softmax Loss for Item Recommendation}.
\newblock \bibinfo{journal}{\emph{ACM Trans. Inf. Syst.}} \bibinfo{volume}{42}, \bibinfo{number}{4}, Article \bibinfo{articleno}{98} (\bibinfo{date}{March} \bibinfo{year}{2024}), \bibinfo{numpages}{26}~pages.
\newblock
\showISSN{1046-8188}
\href{https://doi.org/10.1145/3637061}{doi:\nolinkurl{10.1145/3637061}}


\bibitem[Wu et~al\mbox{.}(2016)]%
        {wordpiece}
\bibfield{author}{\bibinfo{person}{Yonghui Wu}, \bibinfo{person}{Mike Schuster}, \bibinfo{person}{Zhifeng Chen}, \bibinfo{person}{Quoc~V. Le}, \bibinfo{person}{Mohammad Norouzi}, \bibinfo{person}{Wolfgang Macherey}, \bibinfo{person}{Maxim Krikun}, \bibinfo{person}{Yuan Cao}, \bibinfo{person}{Qin Gao}, \bibinfo{person}{Klaus Macherey}, \bibinfo{person}{Jeff Klingner}, \bibinfo{person}{Apurva Shah}, \bibinfo{person}{Melvin Johnson}, \bibinfo{person}{Xiaobing Liu}, \bibinfo{person}{Łukasz Kaiser}, \bibinfo{person}{Stephan Gouws}, \bibinfo{person}{Yoshikiyo Kato}, \bibinfo{person}{Taku Kudo}, \bibinfo{person}{Hideto Kazawa}, \bibinfo{person}{Keith Stevens}, \bibinfo{person}{George Kurian}, \bibinfo{person}{Nishant Patil}, \bibinfo{person}{Wei Wang}, \bibinfo{person}{Cliff Young}, \bibinfo{person}{Jason Smith}, \bibinfo{person}{Jason Riesa}, \bibinfo{person}{Alex Rudnick}, \bibinfo{person}{Oriol Vinyals}, \bibinfo{person}{Greg Corrado}, \bibinfo{person}{Macduff Hughes}, {and} \bibinfo{person}{Jeffrey Dean}.}
  \bibinfo{year}{2016}\natexlab{}.
\newblock \bibinfo{title}{Google's Neural Machine Translation System: Bridging the Gap between Human and Machine Translation}.
\newblock
\showeprint[arxiv]{1609.08144}~[cs.CL]
\urldef\tempurl%
\url{https://arxiv.org/abs/1609.08144}
\showURL{%
\tempurl}


\bibitem[Yan et~al\mbox{.}(2024)]%
        {trinity}
\bibfield{author}{\bibinfo{person}{Jing Yan}, \bibinfo{person}{Liu Jiang}, \bibinfo{person}{Jianfei Cui}, \bibinfo{person}{Zhichen Zhao}, \bibinfo{person}{Xingyan Bin}, \bibinfo{person}{Feng Zhang}, {and} \bibinfo{person}{Zuotao Liu}.} \bibinfo{year}{2024}\natexlab{}.
\newblock \showarticletitle{Trinity: Syncretizing Multi-/Long-Tail/Long-Term Interests All in One}. In \bibinfo{booktitle}{\emph{Proceedings of the 30th ACM SIGKDD Conference on Knowledge Discovery and Data Mining}} (Barcelona, Spain) \emph{(\bibinfo{series}{KDD '24})}. \bibinfo{publisher}{Association for Computing Machinery}, \bibinfo{address}{New York, NY, USA}, \bibinfo{pages}{6095–6104}.
\newblock
\showISBNx{9798400704901}
\href{https://doi.org/10.1145/3637528.3671651}{doi:\nolinkurl{10.1145/3637528.3671651}}


\bibitem[Yang et~al\mbox{.}(2020)]%
        {mns}
\bibfield{author}{\bibinfo{person}{Ji Yang}, \bibinfo{person}{Xinyang Yi}, \bibinfo{person}{Derek Zhiyuan~Cheng}, \bibinfo{person}{Lichan Hong}, \bibinfo{person}{Yang Li}, \bibinfo{person}{Simon Xiaoming~Wang}, \bibinfo{person}{Taibai Xu}, {and} \bibinfo{person}{Ed~H. Chi}.} \bibinfo{year}{2020}\natexlab{}.
\newblock \showarticletitle{Mixed Negative Sampling for Learning Two-tower Neural Networks in Recommendations}. In \bibinfo{booktitle}{\emph{Companion Proceedings of the Web Conference 2020}} (Taipei, Taiwan) \emph{(\bibinfo{series}{WWW '20})}. \bibinfo{publisher}{Association for Computing Machinery}, \bibinfo{address}{New York, NY, USA}, \bibinfo{pages}{441–447}.
\newblock
\showISBNx{9781450370240}
\href{https://doi.org/10.1145/3366424.3386195}{doi:\nolinkurl{10.1145/3366424.3386195}}


\bibitem[Yi et~al\mbox{.}(2019)]%
        {logq}
\bibfield{author}{\bibinfo{person}{Xinyang Yi}, \bibinfo{person}{Ji Yang}, \bibinfo{person}{Lichan Hong}, \bibinfo{person}{Derek~Zhiyuan Cheng}, \bibinfo{person}{Lukasz Heldt}, \bibinfo{person}{Aditee Kumthekar}, \bibinfo{person}{Zhe Zhao}, \bibinfo{person}{Li Wei}, {and} \bibinfo{person}{Ed Chi}.} \bibinfo{year}{2019}\natexlab{}.
\newblock \showarticletitle{Sampling-bias-corrected neural modeling for large corpus item recommendations}. In \bibinfo{booktitle}{\emph{Proceedings of the 13th ACM Conference on Recommender Systems}} (Copenhagen, Denmark) \emph{(\bibinfo{series}{RecSys '19})}. \bibinfo{publisher}{Association for Computing Machinery}, \bibinfo{address}{New York, NY, USA}, \bibinfo{pages}{269–277}.
\newblock
\showISBNx{9781450362436}
\href{https://doi.org/10.1145/3298689.3346996}{doi:\nolinkurl{10.1145/3298689.3346996}}


\end{thebibliography}

\appendix
\onecolumn

\section{Investigating Standard LogQ Correction}

In this section, we provide a derivation of the standard logQ correction loss:
\begin{equation}\label{default_logq}
\mathcal{L}_{\text{logQ}}(u, p) = -\log \frac{e^{f_\theta(u, p) - \log Q(p)}}{e^{f_\theta(u, p) - \log Q(p)} + \sum_{i=1}^n e^{f_\theta(u, d_i) - \log Q(d_i)}}.
\end{equation}
The primary objective of this correction is to approximate the full softmax formulation:
\[
\mathcal{L}_{\text{softmax}}(u, p) = -\log \frac{e^{f_\theta(u, p)}}{\sum_{d \in \mathcal{D}} e^{f_\theta(u, d)}},
\]
where $\mathcal{D}$ is the entire item catalog. 

We first show that \(\nabla_\theta \mathcal{L}_{\text{softmax}}(u, p)\) can be expressed as an expectation over the entire item catalog:
\begin{align*}
    \nabla_\theta \mathcal{L}_{\text{softmax}}(u, p) &=
    \nabla_\theta \left( -\log \frac{e^{f_\theta(u, p)}}{\sum_{d \in \mathcal{D}} e^{f_\theta(u, d)}} \right) = - \nabla_\theta f_\theta(u, p) +\sum\limits_{d \in \mathcal{D}}\underbrace{\frac{e^{f_\theta(u, d)}}{\sum_{\hat d \in \mathcal{D}}e^{f_\theta(u, \hat d)}}}_{P_\theta(d \mid u)}\nabla_\theta f_\theta(u, d) = - \nabla_\theta f_\theta(u, p) + \mathbb{E}_{d \sim P_\theta(d \mid u)} \left[\nabla_\theta f_\theta(u, d) \right].
\end{align*}
We next employ the importance sampling to replace \(P_\theta(d \mid u)\) in the expectation with the proposal distribution \(Q(d)\):
\begin{align*}
    \nabla_\theta \mathcal{L}_{\text{softmax}}(u, p) &= - \nabla_\theta f_\theta(u, p) + \mathbb{E}_{d \sim P_\theta(d \mid u)} \left[\nabla_\theta f_\theta(u, d) \right] \\ 
    &= - \nabla_\theta f_\theta(u, p) + \mathbb{E}_{d \sim Q(d)} \left[\frac{P_\theta(d \mid u)}{Q(d)}\nabla_\theta f_\theta(u, d) \right] \\ 
    & \approx - \nabla_\theta f_\theta(u, p) + \frac{1}{n}\sum\limits_{i=1}^{n} \left( \frac{1}{Q(d_i)}P_\theta(d_i \mid u) \nabla_\theta f_\theta(u, d_i) \right)\\ 
    &= - \nabla_\theta f_\theta(u, p) + \frac{1}{n}\sum\limits_{i=1}^{n} \left( \frac{e^{f_\theta(u, d_i) - \log Q(d_i)}}{\sum_{\hat d \in \mathcal{D}}e^{f_\theta(u, \hat d)}}\nabla_\theta f_\theta(u, d_i) \right),
\end{align*}
where $d_i \sim Q(d)$. Subsequently, we approximate the denominator term using the same importance sampling technique:
\begin{align}\label{for_consistency}
    \sum\limits_{ d \in \mathcal{D}}e^{f_\theta(u,  d)} &= |\mathcal{D}| \cdot \mathbb{E}_{ d \sim \text{Unif($\mathcal{D}$)}} \left[e^{f_\theta(u,  d)}\right] = |\mathcal{D}| \cdot \mathbb{E}_{ d \sim Q( d)} \left[\frac{1}{|\mathcal{D}|}e^{f_\theta(u,  d) - \log Q( d)}\right] \approx \frac{1}{n}\sum\limits_{k=1}^{n}e^{f_\theta(u,  d_k) - \log Q( d_k)}.
\end{align}
Substituting this approximation into our gradient expression yields:
\begin{align*}
    \nabla_\theta \mathcal{L}_{\text{softmax}}(u, p) &\approx - \nabla_\theta f_\theta(u, p) + \frac{1}{n}\sum\limits_{i=1}^{n} \left( \frac{e^{f_\theta(u, d_i) - \log Q(d_i)}}{\frac{1}{n}\sum_{k=1}^{n}e^{f_\theta(u,  d_k) - \log Q( d_k)}}\nabla_\theta f_\theta(u, d_i) \right) \\
    &= - \nabla_\theta f_\theta(u, p) + \sum\limits_{i=1}^{n} \left( \frac{e^{f_\theta(u, d_i) - \log Q(d_i)}}{\sum_{k=1}^{n}e^{f_\theta(u, d_k) - \log Q(d_k)}}\nabla_\theta f_\theta(u, d_i) \right)\\
    &= \nabla_\theta \mathcal{L}_{\text{original}}(u, p),
\end{align*}
where $\mathcal{L}_{\text{original}}(u, p) := -\log \dfrac{e^{f_\theta (u, p)}}{\sum_{i=1}^n e^{f_\theta (u, d_i) - \log Q(d_i)}}$.

\paragraph{Weighted importance sampling.} A simpler way to derive the same estimate is to use weighted importance sampling:
\begin{align*}
    \nabla_\theta \mathcal{L}_{\text{softmax}}(u, p) &= - \nabla_\theta f_\theta(u, p) + \mathbb{E}_{d \sim Q(d)} \left[\frac{P_\theta(d \mid u)}{Q(d)}\nabla_\theta f_\theta(u, d) \right] \\ 
    & \approx - \nabla_\theta f_\theta(u, p) + \dfrac{\sum_{i=1}^{n} \frac{P_\theta(d_i \mid u)}{Q(d_i)} \nabla_\theta f_\theta(u, d_i)}{\sum_{i=1}^{n} \frac{P_\theta(d_i \mid u)}{Q(d_i)} } \\
    &= - \nabla_\theta f_\theta(u, p) + \sum\limits_{i=1}^{n} \left( \frac{e^{f_\theta(u, d_i) - \log Q(d_i)}}{\sum_{k=1}^{n}e^{f_\theta(u, d_k) - \log Q(d_k)}}\nabla_\theta f_\theta(u, d_i) \right) \\
    &= \nabla_\theta \mathcal{L}_{\text{original}}(u, p),
\end{align*}
where $d_i \sim Q(d)$. Since weighted importance sampling is a consistent estimator, the resulting gradient estimate is also consistent, i.e. $\nabla_\theta \mathcal{L}_{\text{original}}(u, p)$ converges in probability to the true value $\nabla_\theta \mathcal{L}_{\text{softmax}}(u, p)$ as $n\to\infty$.

\paragraph{Role of the positive item in logQ correction.} Note that one can incorporate \(-\log Q(p)\) into the numerator without affecting the optimization process, as it represents an additive constant:
\[
-\log \frac{e^{f_\theta (u, p) - \log Q(p)}}{\sum_{i=1}^n e^{f_\theta (u, d_i) - \log Q(d_i)}} = -\log \left(\frac{1}{Q(p)} \cdot \frac{e^{f_\theta (u, p)}}{\sum_{i=1}^n e^{f_\theta (u, d_i) - \log Q(d_i)}} \right) = \mathcal{L}_{\text{original}}(u, p) + \log Q(p) = \mathcal{L}_{\text{original}}(u, p) + \text{const}.
\]

In practice, the derived logQ correction is frequently combined with the inclusion of the positive example in the loss denominator,
as in formula (\ref{default_logq}). This approach better aligns with the inference-time ranking task by placing positives and negatives within the same normalization context. However, it is important to recognize that including positive in the loss denominator introduces a theoretical inconsistency with our derivation, as we assume negatives are sampled from a proposal distribution \(Q\), whereas the positive is included deterministically with probability 1. 

Nevertheless, the gradient $\nabla_\theta\mathcal{L}_{\text{logQ}}(u, p)$ also converges in probability to $\nabla_\theta\mathcal{L}_{\text{softmax}}(u, p)$ as $n\to\infty$. To show this, let us introduce the notation
$$
v_{up} := \frac{e^{f_\theta(u, p) - \log Q(p)}}{e^{f_\theta(u, p) - \log Q(p)} + \sum_{i=1}^n e^{f_\theta(u, d_i) - \log Q(d_i)}}.
$$
The following identity is straightforward for each $1\le k \le n$:
$$
\frac{1}{e^{f_\theta(u, p) - \log Q(p)} + \sum_{i=1}^n e^{f_\theta(u, d_i) - \log Q(d_i)}} = \frac{1 - v_{up}}{ \sum_{i=1}^n e^{f_\theta(u, d_i) - \log Q(d_i)}}.
$$
Using it, we obtain the formula for the gradient $\nabla_\theta\mathcal{L}_{\text{logQ}}(u, p)$:
\begin{align*}
\nabla_\theta\mathcal{L}_{\text{logQ}}(u, p) &= - \nabla_\theta \log v_{up} \\ 
&= - \nabla_\theta f_\theta(u, p) + \nabla_\theta\log\left(e^{f_\theta(u, p) - \log Q(p)} + \sum_{i=1}^n e^{f_\theta(u, d_i) - \log Q(d_i)}\right) \\
&= - \nabla_\theta f_\theta(u, p) + \frac{e^{f_\theta(u, p) - \log Q(p)}\nabla_\theta f_\theta(u, p) + \sum_{i=1}^n e^{f_\theta(u, d_i)- \log Q(d_i)}\nabla_\theta f_\theta(u, d_i)}{e^{f_\theta(u, p) - \log Q(p)} + \sum_{i=1}^n e^{f_\theta(u, d_i) - \log Q(d_i)}} \\
&= - \nabla_\theta f_\theta(u, p) + v_{up} \nabla_\theta f_\theta(u, p) + (1-v_{up})
\frac{\sum_{i=1}^n e^{f_\theta(u, d_i) - \log Q(d_i)}\nabla_\theta f_\theta(u, d_i)}{ \sum_{i=1}^n e^{f_\theta(u, d_i) - \log Q(d_i)}}\\
&= (1-v_{up})\left(-\nabla_\theta f_\theta(u, p) + \sum_{i=1}^n \frac{e^{f_\theta(u, d_i) - \log Q(d_i)}}{ \sum_{k=1}^n e^{f_\theta(u, d_k) - \log Q(d_k)}}\nabla_\theta f_\theta(u, d_i)\right) \\
&= (1-v_{up})\nabla_\theta\mathcal{L}_{\text{original}}(u, p).
\end{align*}
From (\ref{for_consistency}), we obtain
$$
1 - v_{up} = \frac{\sum_{i=1}^n e^{f_\theta(u, d_i) - \log Q(d_i)}}{e^{f_\theta(u, p) - \log Q(p)} + \sum_{i=1}^n e^{f_\theta(u, d_i) - \log Q(d_i)}} = \frac{\frac1n \sum_{i=1}^n e^{f_\theta(u, d_i) - \log Q(d_i)}}{\frac1n e^{f_\theta(u, p) - \log Q(p)} + \frac1n \sum_{i=1}^n e^{f_\theta(u, d_i) - \log Q(d_i)}} \xrightarrow{P} \frac{\sum_{ d \in \mathcal{D}}e^{f_\theta(u,  d)}}{\sum_{ d \in \mathcal{D}}e^{f_\theta(u,  d)}} = 1~\text{as}~n\to\infty.
$$
Hence, $\nabla_\theta\mathcal{L}_{\text{logQ}}(u, p)\xrightarrow{P} \nabla_\theta\mathcal{L}_{\text{softmax}}(u, p)$ as $n\to\infty$. The same derivation holds for the loss with standard logQ correction, but without the correction applied to the positive example in the denominator:
$$
 -\log \frac{e^{f_\theta(u, p)}}{e^{f_\theta(u, p)} + \sum_{i=1}^n e^{f_\theta(u, d_i) - \log Q(d_i)}}.
$$

\section{Correcting the Correction: Full Derivation}

Following the derivation of the standard logQ correction, we first approximate the full softmax gradient. Unlike the standard derivation, our approach explicitly models the positive item \(p\) in the denominator, thus resolving the issue described above:

\begin{align*} 
\nabla_\theta \mathcal{L}_{\text{softmax}}(u, p) &= \nabla_\theta \left[ -f_\theta(u, p) + \log \sum_{d \in \mathcal{D}} e^{f_\theta(u, d)} \right] \\
&= -\nabla_\theta f_\theta(u, p) + \sum_{d \in \mathcal{D}} \frac{e^{f_\theta(u, d)}}{\sum_{\hat d \in \mathcal{D}} e^{f_\theta(u, \hat d)}} \nabla_\theta f_\theta(u, d)\\ 
&= -\nabla_\theta f_\theta(u, p) + \sum_{d \in \mathcal{D}} P_\theta(d \mid u) \nabla_\theta f_\theta(u, d) \\ 
&= -(1 - P_\theta(p \mid u)) \nabla_\theta f_\theta(u, p) + \sum_{d \in \mathcal{D} \setminus \{p\}} P_\theta(d \mid u) \nabla_\theta f_\theta(u, d) \\
&= -(1 - P_\theta(p \mid u)) \nabla_\theta f_\theta(u, p) + \sum_{d \in \mathcal{D} \setminus \{p\}} \underbrace{\frac{e^{f_\theta(u, d)}}{\sum_{\hat d \in \mathcal{D} \setminus \{p\}}e^{f_\theta(u, \hat d)}}}_{P_\theta(d \mid u, d \neq p)} \underbrace{\frac{\sum_{\hat d \in \mathcal{D} \setminus \{p\}}e^{f_\theta(u, \hat d)}}{\sum_{\hat d \in \mathcal{D}} e^{f_\theta(u, \hat d)}}}_{(1 - P_\theta(p \mid u))} \nabla_\theta f_\theta(u, d) \\
&= (1 - P_\theta(p \mid u)) \left( -\nabla_\theta f_\theta(u, p) + \sum_{d \in \mathcal{D} \setminus \{p\}} P_\theta(d \mid u, d \neq p) \nabla_\theta f_\theta(u, d) \right) \\
&= (1 - P_\theta(p \mid u)) \left( -\nabla_\theta f_\theta(u, p) + \mathbb{E}_{d \sim P_\theta(d \mid u, d \neq p)} \left[ \nabla_\theta f_\theta(u, d) \right] \right).
\end{align*}

We next employ weighted importance sampling to replace \(P_\theta(d \mid u, d \neq p)\) distribution with the proposal \(Q'(d)\) over \(\mathcal{D} \setminus \{p\}\) which is different from \(Q(d)\) in that it explicitly excludes the positive item  \(p\) from being sampled: 
\begin{align*}
    \nabla_\theta \mathcal{L}_{\text{softmax}}(u, p) &= (1 - P_\theta(p \mid u)) \left( -\nabla_\theta f_\theta(u, p) + \mathbb{E}_{d \sim P_\theta(d \mid u, d \neq p)} \left[ \nabla_\theta f_\theta(u, d) \right] \right) \\ 
    &= (1 - P_\theta(p \mid u)) \left( -\nabla_\theta f_\theta(u, p) + \mathbb{E}_{d \sim Q'(d)} \left[ \frac{P_\theta(d \mid u, d \neq p)}{Q'(d)} \nabla_\theta f_\theta(u, d) \right] \right) \\
    &= (1 - P_\theta(p \mid u)) \left( -\nabla_\theta f_\theta(u, p) + \frac{\sum_{i=1}^{n}\frac{P_\theta(d_i \mid u, d_i \neq p)}{Q'(d_i)} \nabla_\theta f_\theta(u, d_i)}{\sum_{i=1}^{n}\frac{P_\theta(d_i \mid u, d_i \neq p)}{Q'(d_i)}} \right)  \\
    &= (1 - P_\theta(p \mid u)) \left( -\nabla_\theta f_\theta(u, p) + \sum_{i=1}^{n} \frac{e^{f_\theta(u, d_i) - \log Q'(d_i)}}{\sum_{k=1}^{n}e^{f_\theta(u, d_k) - \log Q'(d_k)}}\nabla_\theta f_\theta(u, d_i) \right),
\end{align*}
where ${d_1, \ldots, d_n}$ are sampled from $Q'(d)$ -- proposal distribution over $\mathcal{D} \setminus \{p\}$.

The resulting gradient estimate corresponds to the following loss function:
\[
\mathcal{L}_{\text{ours}}(u, p) = - \text{sg}(1 - P_\theta(p \mid u)) \log \frac{e^{f_\theta (u, p)}}{\sum_{i=1}^n e^{f_\theta (u, d_i) - \log Q'(d_i)}},
\]
where $\text{sg}$ denotes the stop-gradient operation (no gradient flows through \(P_\theta(p \mid u)\)). To estimate $P_\theta(p \mid u)$, importance sampling is applied once more:
\begin{align*} 
P_\theta(p \mid u) 
&= \frac{e^{f_\theta(u, p)}}{\sum_{d \in \mathcal{D}} e^{f_\theta(u, d)}} \\
&= \frac{e^{f_\theta(u, p)}}{e^{f_\theta(u, p)} + \sum_{d \in \mathcal{D} \setminus \{p\}} e^{f_\theta(u, d)}} \\ 
&= \frac{e^{f_\theta(u, p)}}{e^{f_\theta(u, p)} + (|\mathcal{D}|-1) \mathbb{E}_{d \sim \text{Unif}(\mathcal{D} \setminus \{p\})} \left[ e^{f\theta(u, d)} \right]} \\ 
&\approx \frac{e^{f_\theta(u, p)}}{e^{f_\theta(u, p)} + \frac{1}{n} \sum_{i=1}^n e^{f_\theta(u, d_i) - \log Q'(d_i)}}, \quad d_i \sim Q'(d). 
\end{align*}

Estimates for both $P_\theta(p \mid u)$ and $\mathbb{E}_{d \sim P_\theta(d \mid u, d \neq p)} \left[ \nabla_\theta f_\theta(u, d) \right]$ are consistent, thus the same is true for the overall estimate, i.e. $\nabla_\theta\mathcal{L}_{\text{ours}}(u, p)$ converges in probability to $\nabla_\theta\mathcal{L}_{\text{softmax}}(u, p)$ as $n\to\infty$.


\end{document}